# Research on solitons' interactions in one-dimensional indium chains on Si(111) surfaces

Y Yao[1,3], C J Luo[1], X X Wang[1,4], H Zhang[1,2,3*]

[1] Hefei National Research Center for Physical Sciences at the Microscale, University of Science and Technology of China, Hefei, Anhui, 230026, China
[2] Hefei National Laboratory, University of Science and Technology of China, Hefei, Anhui, 230026, China
[3] Department of Physics, University of Science and Technology of China, Hefei, Anhui, 230026, China
[4] Center for Micro- and Nanoscale Research and Fabrication, University of Science and Technology of China, Hefei, Anhui, 230026, China

*Email: huiz@ustc.edu.cn

**Abstract**. Solitons have garnered significant attention across various fields, yet a contentious debate persists regarding the precise structure of solitons on indium chains. Currently, multiple forms of solitons in one-dimensional atomic chains have been reported. STM provides an effective means to study the precise atomic structure of solitons, particularly their dynamics and interactions. However, limited research has been conducted on soliton interactions and soliton-chain interactions, despite their profound impact on relative soliton motions and the overall physical properties of the system. In this work, we characterized the structures of the soliton dimer and trimer, observed the displacements induced by the soliton entity and statisticized the dynamic behaviors of soliton dimers over time evolution or temperature. To reveal the soliton mechanism, we further utilized STM to investigate the CDWs between two solitons when two monomers were encountered. Additionally, we achieved the manipulation of the monomer on the indium chain by the STM tip. Our work serves as an important approach to elucidate interactions in correlated electronic systems and advance the development of potential topological soliton computers.

## 1. Introduction

One-dimensional (1D) materials have aroused intensive interest because of their exotic physical properties and potential applications in nanoelectronics. The interactions significantly enhanced between spin, charge, and lattice in 1D electronic systems can result in singular ground states and excited states [1,2,3] as well as charge density waves (CDWs) [4]. Reportedly, some metals on certain semiconductor surfaces at low coverages would be self-organized to generate atomic chains [5] and exhibit one-dimensional charge density wave states at low temperatures [6]. In addition to the presence of exotic ground states, the elementary excitations of one-dimensional CDWs are even more intriguing. The nonlinear topological excitation or soliton [7] has been well studied in 1D conjugated polymers such as polyacetylene [8,9] and some preliminary indications of soliton behaviors at the atomic scale have also been recently reported by STM [10,11,12]. Nevertheless, there still remains a debate on the precise structure of solitons. At present, it is reported that there are four forms of solitons in 1D atomic chains: domain wall solitons [13], chiral solitons [14], fractional solitons [15] and topological solitons

[16,17,18]. Compared with other solitons, topological solitons have a well-defined structure, formation energy and dynamic energy at the atomic scale. Soliton movement, annihilation, creation and manipulation can be observed by STM [18]. However, as there might be many solitons crowding on the 1D indium chain, the soliton-soliton interaction and soliton-chain interaction will greatly influence the relative soliton motions and therefore play a vital role in the whole system's physical properties.

In this work, we characterized the structures of the soliton dimer and trimer, observed the displacements induced by the soliton entity and statisticized the dynamic behaviors of soliton dimers over time evolution or temperature. To reveal the soliton mechanism, we further utilized STM to investigate the CDWs between two solitons when two monomers were encountered. Additionally, we achieved the manipulation of the monomer on the indium chain by the STM tip.

## 2. Experimental Methodology and Analysis

Our experiments were performed in an ultrahigh vacuum (UHV) system equipped with an Omicron low temperature (LT) STM, with a base pressure below $1 \times 10^{-10}$ mbar. The Si(111) substrates (As doped, $n$ type, ~0.001 Ω·cm) were cleaned by means of the standard "flashing" recipe until the $7 \times 7$ reconstruction was established at the surface. Then, we prepared the Si(111)-($4 \times 1$)-In atomic chain structure by evaporating approximately one monolayer of In onto the cleaned Si(111) surface at 700 K, followed by a post annealing process at the same temperature for 30 minutes.

### 2.1. The Structures of Soliton Dimer and Trimer

Topological soliton monomers exist on the surface, as evidenced by two $4 \times 2$ chains with alternating hexagon orientations, producing a domain boundary [18]. The structure of the soliton monomer exhibits a heart-shaped phase boundary and connects two mirror-symmetric indium hexagons. Except for soliton monomers, we found soliton dimers and trimers aggregated by monomers, as shown in figure 1(a) and (c). Notably, as the CDW changes according to different bias voltages, the dark chain area represents the indium chain in our experiments. Through STM images, the structure of solitons is totally different from the pair of solitons. In the case of the pair of solitons, one soliton links another soliton with the same phase boundary, while in the case of the dimer and trimer, two or three soliton monomers are connected with different phase boundaries. As illustrated in figure 1(b), as soon as two monomers form a dimer, the phase and boundary shift accordingly. It is also observed that the A domain turns in the A* direction on the soliton dimer (the electronic state of the A* domain is the same as that of the A domain, while their phases are completely different).

The structures of the dimer and trimer are measured. The indium hexagonal length of one soliton is determined to be $3a$ and $2a$ ($a$ is the projection of atomic spacing between vertical chains) at two neighboring indium chains. In the experiment, when two solitons get closer and form a dimer, the distance between the outermost two indium hexagonal lengths becomes $5a$. At this time, the central distance between two solitons is $2.5a$, as shown in figure 1(c). The trimer's structure is formed by combining three solitons, and the distance between the outermost two In hexagons is $7a/8a$. For a higher solitary poly-structure, the distance between two In hexagons is $10a/10a$. Given that the distance between the outermost two In hexagonal lengths is odd $a$ rather than even $a$, there are no soliton annihilation cases. According to previous research [18], the effective mass of the soliton monomer is approximately 3700 electron rest mass. As the number of dimer atoms is almost double that of the monomer, the weight of the dimer is close to double that of the monomer.

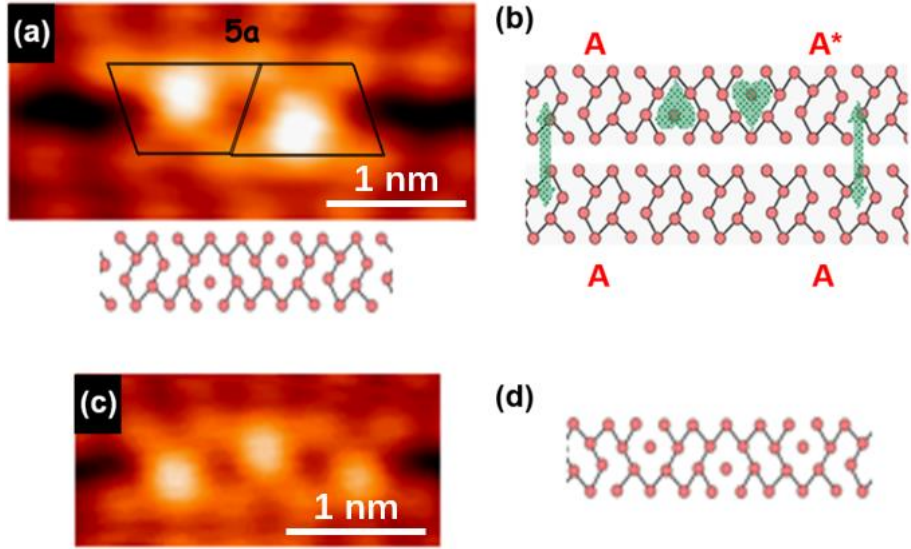

**Figure 1.** Typical STM images of In chains on a Si(111) surface at 78 K at sample bias ($V_s$) of 1.5 V and time resolution of 1 s. (a) and (c) depict the STM images of the soliton dimer and trimer respectively. The ladder-shaped inset of (a) represents the outline of single soliton. (b) and (d) illustrate the atomic structural diagrams of the soliton dimer and trimer, respectively.

*2.2. The Dynamic Properties of Soliton Dimer*

Soliton monomers and dimers are statisticized to explore the soliton system's dynamic equilibrium. The detailed statistical method is similar to the previously reported method for counting soliton monomers [18]. Figure 2(c) and (d) contain ~500 solitons (including monomers and dimers) counted at 78 K, and figure 2(a) and (b) present STM images with a time interval of 70 minutes in the same region. Here, we employ the displacement ratio dN/Ndt to show the possibility of soliton movements, which was obtained by calculating the fraction of the solitons that has changed their positions between two consecutive STM images. Notably, over ~85% monomers moved while most dimers remained stationary. After 70 minutes, the probability that a single soliton does not move was ~80%, while the probability of a dimer was only ~50%. We attributed this phenomenon to the different effective masses of the monomer and dimer, as a larger effective mass will slow the soliton movement and weaken the combination to form dimers or trimers. However, it is also observed that the total number of dimers and trimers remains unchanged after 70 minutes, with the ratio of solitons in both directions always remaining at 1∶1, which suggests that despite soliton annihilation in different directions and polymer formation, the overall number remains conserved.

In addition, the temperature-dependent soliton density, including monomers, dimers and trimers, was investigated to explore the thermal properties. As the temperature increased, the number of dimers and trimers increased correspondingly, and the number of single solitons decreased. To clarify this phenomenon, we hold the opinion that as the temperature climbs, the soliton system is affected by a larger thermal perturbation, making it easier for single solitons to combine with each other and form dimers and pushing the thermal equilibrium between dimer formation and breaking toward dimer formation. To verify the assumption that the soliton aggregation is a temperature-dependent relationship, we decreased the sample temperature to 77 K and then the probability of soliton dimers and trimers decomposed. Notably, when the temperature decreased to lower than 60 K (below the generation energy temperature), we did not observe any monomers, dimers or trimers. When it was heated back to the original temperature, we could hardly observe soliton monomers. The reason why it is difficult to find soliton monomers on the surface at high temperature is explained below.

As solitons are only found at the 8 × 2 phase [18], some soliton onsite motions do not satisfy the characteristics of one-dimensional random walking. We performed relevant experiments to demonstrate CDW transitions. Due to the short time interval between each scan line of STM, we cannot acquire the entire image of solitons; therefore, only one rapid phase transition was observed on the indium chain. In a one-dimensional system, even a small onsite perturbation can result in soliton movement and CDW segment phase transition, indicating collective behavior in a one-dimensional atomic chain.

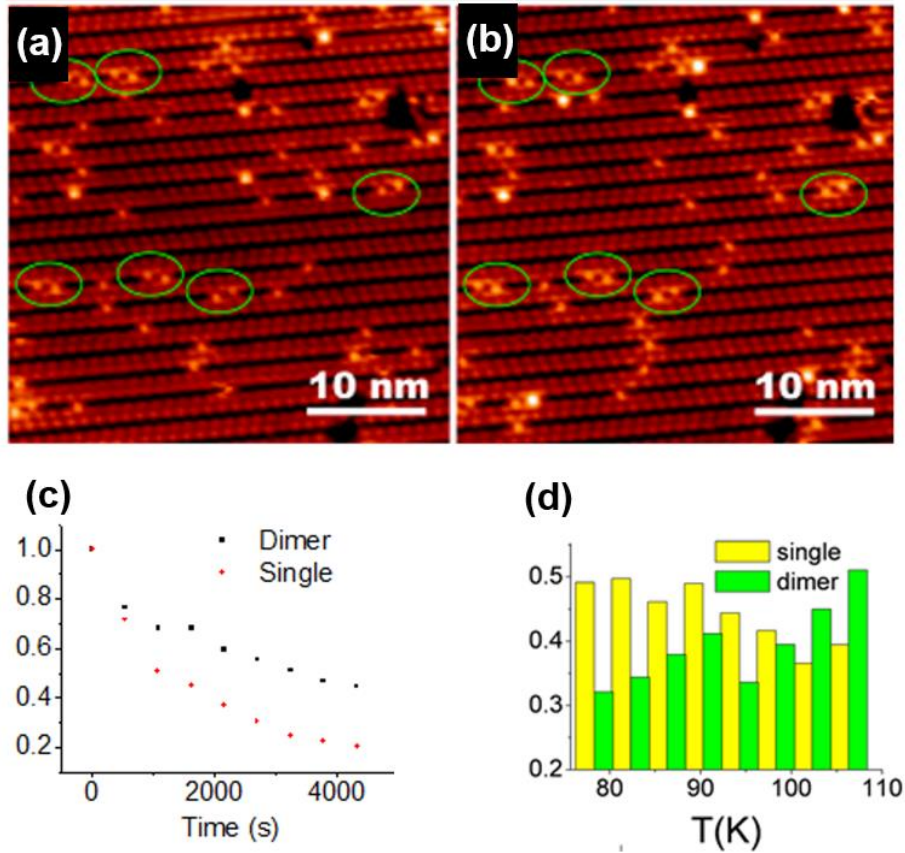

**Figure 2.** (a) (b) STM images taken with a 70-minute interval in the same region at time resolution of 1 s. The green circles indicate relatively stable soliton dimers. (c) Possibility diagram illustrating the density of stationary single solitons versus the density of stationary soliton dimers over time, obtained from the temperature variation plot at 86 K. (d) Temperature-dependent possibility diagram illustrating the stationary soliton monomers versus the stationary soliton dimers.

*2.3. The Relationship between Soliton Dynamic Behavior and CDW*

To understand the relationship between the soliton dynamic behavior and its CDW, we characterized two encountering solitons with STM. In the experiment, initially, two solitons at the same indium zigzag chain stay far away from each other and then they begin to get closer to each other. When the distance between them is lower than $10a$, the wavelength of the charge density waves between two solitons will start to change, as shown in figure 3(d). Notably, the distance between two indium hexagons at one line is $9a$, the distance at another line is $8a$, and a non-integer periodicity of the charge density wave exists between the two indium hexagons. It is difficult to explain that between two indium hexagons at a distance of $8a$, the CDW is not consistent with the periodic CDW at multiplate $2a$, instead of the same periodic CDW resembling chains of $9a$. Through experimental evidence, the distribution of the charge density is determined to be centered on the midpoint between two solitons, as a result of which two solitons form a stable structure on the indium chain.

We also witnessed the phase inversion of the CDW between two encountering solitons. When the distance between two solitons becomes longer than $12a$, the interaction weakens, and the periodic CDW switches back to the normal case. Figure 3(b) and (c) show the CDW transitioning between two solitons. Through STM images, the distance between two indium hexagons is determined to be $7a/6a$ at different indium lines. After 60 minutes, the distance between two In hexagons at the upper and lower Indium zigzag lines decreases following $9a/8a$, $7a/6a$, $5a/4a$ and $3a/2a$ (the closest state between two solitons). According to the length of the indium zigzag chain, the distance at both chains consists of the trimer solitons. As two solitons continue to get closer, their interactions will increase drastically, which will induce one of the solitons to shift its direction and then form a stable dimer structure. Figure 3(e) and (f) demonstrate the process of how two solitons in the same chain combine into a soliton dimer. Figure 3(f) shows the comparison between solitons, where the right soliton still maintains its original orientation and morphology while the left soliton reverses its direction.

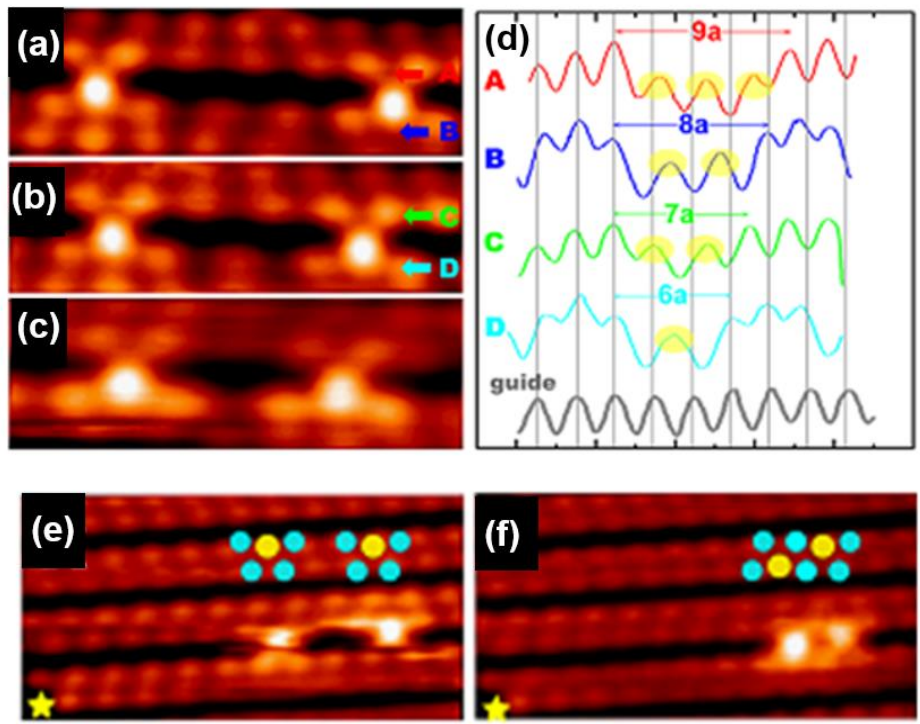

**Figure 3.** (a)-(c) STM images demonstrating the encounter of two solitons at the same Indium chain at time resolution of 1 s. (d) The variation in chain length between the two solitons. (e) (f) Two solitons with the same direction meeting and interacting, resulting in the transformation into two solitons with opposite directions.

*2.4. The Manipulation on Soliton Monomer by STM Tip*

In addition to soliton aggregation, dimers and trimers can also be dissolved to form monomers. The electric field from the STM tip will induce Peierls instability on the chain segment and lead to soliton movement, which will promote soliton dimer and trimer formation and breaking. However, the aggregation of solitons to form stable dimer and trimer structures is effectively weakened. Compared to traditional systems, statistically, the number of solitons attracted to each other is more than the number of solitons repulsed to each other. On account of the dimer's slow motion and its transparent dependence on temperature, the soliton monomer eventually becomes a dimer to stabilize the whole indium chain system. Moreover, to compensate for such changes, the system needs to generate or annihilate new solitons to maintain the balance of the total number during the process of generation and annihilation. However, as the temperature increases, the formation energy of solitons increases, and the area of the 8 × 2 region decreases, which increases the interaction between solitons. The increased interaction will also be balanced by the generation of dimers. Because the stable dimer reduces the system's energy, the whole system tends to be stable.

For a further understanding of the interaction between the soliton and the indium chain, we manipulated the soliton displacement by applying a voltage pulse onto the chain. Figure 4(a)-(c) exhibit a series of STM images taken consequentially. The pulse ($V_s$ = 3.5 V) with a period of 3.6 seconds is applied at the position (approximately 2 nm away from the soliton in the same chain) marked by the filled circle in figure 4(b). Within the duration of the pulse, the soliton distinctly shifts towards the blue dot, a fact corroborated by the points observed along the scanning line which progresses from bottom to top. Upon reaching the arrow, as depicted in figures 4 (b), a pulse is administered to the blue dot. It is at this juncture that the soliton undergoes a positional shift. Following the completion of the pulse, the scanning process persists, with the subsequently captured images beyond the arrow illustrating the trajectory of the soliton's movement. After the pulse, soliton continues to move for a short distance, as indicated by the discontinuity of the scanning lines around the pulse position. Figure 4(a) and (c) are the STM images before and after the bias pulse. The soliton diffuses to the left by $4a$ away from the original position. We also discovered that the distance between the tip-induced point and soliton is surprisingly larger than 20 nm, indicating interactions between solitons and the indium chain, which may be attributed to the role of the electric field.

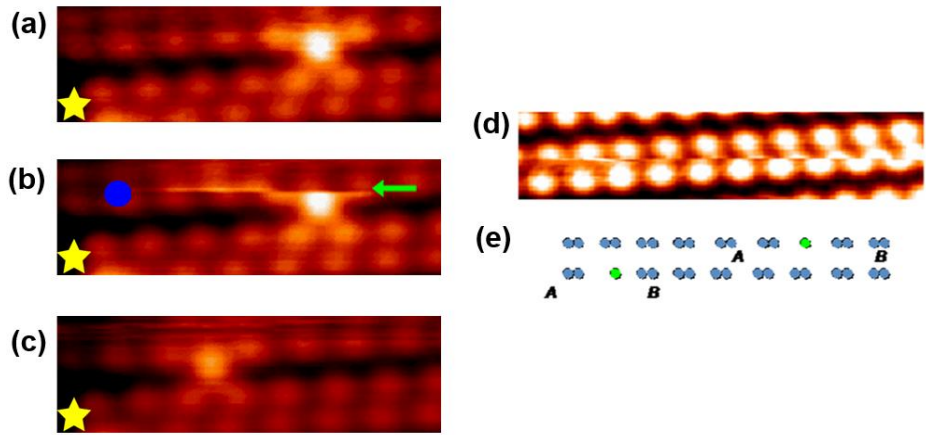

**Figure 4.** (a)-(c) A series of STM images showcases the movement of solitons at time resolution of 1 s. Defects at pentagonal stars can serve as markers. The time interval between the three graphs is 6 minutes, and the scanning bias is 1.5 V. (a) represents the initial state. (b) showcases the process of applying bias on the tip. The blue dot indicates the location where the tip is biased, and the green arrow represents the direction of tip movement. (c) demonstrates the soliton after the pulse. (d)The STM figure illustrating the scenario when a soliton rapidly traverses the In chain. (e) The schematic diagram depicting the whole process of change.

## 3. Conclusion
In conclusion, we observed soliton monomers, dimers and trimers in the one-dimensional indium chain on the Si(111) surface with STM. The soliton dimer and trimer are mainly composed of soliton monomers aggregated in different chains. We statisticized the monomer and dimer densities and investigated the whole process of dimer formation with its CDW segment, concluding that soliton aggregation to form a stable dimer structure will effectively reduce the whole system's energy. Moreover, by applying an electric field between the STM tip and the surface, we achieved atomic manipulation of single solitons. Our research exhibits a new atomic perspective for charge density waves, serves as an important approach to elucidate interactions in correlated electronic systems and may provide a platform for potential topo-logical soliton computers.

**Acknowledgement**
This research was also supported by The National Natural Science Foundation of China (Grants No. 12074357), The Innovation Program for Quantum Science and Technology (No. 2021ZD0302800), The Fundamental Research Funds for the Central Universities (No. WK9990000118, WK2310000104).